\documentclass[twocolumn,english]{IEEEtran}
\usepackage[T1]{fontenc}
\usepackage{amstext}
\usepackage{graphicx}

\makeatletter
\let\oldforeign@language\foreign@language
\DeclareRobustCommand{\foreign@language}[1]{%
  \lowercase{\oldforeign@language{#1}}}

\makeatother

\usepackage{babel}

\begin{document}
\newcommand{\eqref}[1]{(\ref{#1})}

\title{Multi-Amdahl: Optimal Resource Sharing with Multiple Program Execution
Segments}

\author{Tsahee Zidenberg, Isaac Keslassy, and Uri Weiser%
\thanks{All writers are with the department of Electrical Engineering, Technion,
Haifa 32000, Israel. As usual, this technical report has not been
peer-reviewed and does not count as a publication. It simply presents
a few preliminary results.%
}}

\markboth{TECHNICAL REPORT TR11-03, COMNET, TECHNION, ISRAEL}{}
\maketitle
\begin{abstract}
This paper presents Multi-Amdahl, a resource allocation analytical
tool for heterogeneous systems. Our model includes multiple program
execution segments, where each one is accelerated by a specific hardware
unit. The acceleration speedup of the specific hardware unit is a
function of a limited resource, such as the unit area, power, or energy. 

Using the Lagrange theorem we discover the optimal resource distribution
between all specific units. We then illustrate this general Multi-Amdahl
technique using several examples of area and power allocation among
several cores and accelerators.
\end{abstract}

\section{Introduction}

\PARstart{I}{n} the past few years, chip designers have increasingly
taken into account resource constraints, most notably power, as a
design goal. Their focus has shifted from improving performance to
improving performance within a limited power envelope. Heterogeneous
cores have been suggested for performance/power ratio improvement.
These units are designed for specific workloads, trading efficiency
for flexibility. The shift towards special-purpose hardware can be
seen in today's CPU products, which add a graphic accelerator to the
general-purpose cores \cite{SandyBridge,Introductiont,AMDFusionFam}.
Another example of this trend can be seen inside the general-purpose
core; special-purpose logic is added, supporting specific computation,
such as CRC or cryptography \cite{Debunkingthe}. 

In multicore environment it is necessary to correctly balance the
performance of parallel and serial code segments to overcome the Amdahl
law ceiling \cite{AmdahlsLaw}. Parallel code runs most efficiently
when splitting the available area into many processors, while the
serial code can only run on a single processor. The difference between
the requirements of the two sections created the \textit{asymmetric-cores}
approach \cite{ACCMP-Asymme}. An optimal point for the asymmetry
can be found using Amdahl's law for balancing the importance of parallel
and serial execution, along with the fact that all processors share
a common resource \cite{hill2008amdahl}. 

As hardware becomes more specialized and diverse, Amdahl's law becomes
\textit{multidimensional}. In this environment not only some segments
are accelerated while others are not, but also different accelerated
segments might rely on different types of accelerators. Some of these
segments could represent high-level computation sections, such as
matrix multiplication and FFT, while others could represent low-level
computation sections, such as sections with many floating-point instructions.

A limited resource, such as power or area, is shared among the various
specific units on the chip. Our target is to find the optimal way
to distribute that resource between the specific HW units, balancing
the efficiency of different hardware units with their importance and
performance.

This paper presents two main contributions: 
\begin{itemize}
\item We propose Multi-Amdahl, an analytical tool to optimize resource allocation
among $n$ different specific HW units running $n$ segments of execution
code. The architect may impose constraints on the design, such as
total area, power, or energy, and expect optimal outcome, such as
maximum speedup. In our model, we take into account the differences
in efficiency and scalability of hardware units, and the workload
distribution among the different segments. 
\item Initial results and intuitions obtained from Multi-Amdahl are presented.
These results suggest that the opportunities that exist in heterogeneity
might surpass the cost of inflexibility. In other words, the occasional
use of an accelerator might exceed the costs resulting from its frequent
inactivity.
\end{itemize}
\textbf{The paper structure is as follows:} Section II covers the
related work, Section III presents the Multi-Amdahl technique. Our
technique is compared with Amdahl's law in Section IV. In Sections
V and VI, we present a few examples for extending the basic model.
Section VII presents results from our model and their implications,
and Section VIII concludes and points to potential future extensions
of our work.

\section{Related work}

\subsection*{The move towards accelerators}

Venkatesh et al. \cite{ConservationC} explored the problem of {}``dark
silicon''. Today, threshold voltage no longer scales when technology
advances. The result is that a shrinking percent of the chip may be
activated simultaneously. To achieve peak performance, we must make
sure that minimal power is spent for each function, so that more functions
could be executed in parallel. The article suggested an architecture
with many heterogeneous cores, each designed for optimal power efficiency
of a different software function. Those automatically-generated units
provided up to 30x efficiency (work/J) over general-purpose MIPS.
The article also suggested {}``patchable'' versions of those units
with lower (16x) improvements.

Shee et al. \cite{Shee08} tested a few architectures for a heterogeneous
chip built especially to encode JPG images. Each core was optimized
for one specific pipeline stage. Mostly, optimization was done by
removing unnecessary components from a general-purpose core. Hameed
et al. \cite{Understanding-1} created an even more specific processor
for video encoding by adding custom-made {}``magic'' instructions.
The final chip was about $256\times$ faster than the original RISC
processor, while consuming about 1\% of the energy and 126\% of the
area.

Chung et al. \cite{Single-ChipHe} compared the power efficiency of
general-purpose cores with three forms of {}``unconventional cores'':
FPGA, GPGPU and custom logic. Depending on the benchmark, custom logic
was shown to be around 100x times as power-efficient (performance
ratio / power ratio) as a CPU. The ratio for FPGAs and GPGPUs was,
depending on the benchmark, around 10x and 5x respectively.

A large body of work by both academia and industry is dedicated to
supporting the heterogeneous compute environment. Various frameworks
have been suggested for different aspects of the heterogeneous environment,
including programing \cite{OpenCL-slides}, run-time \cite{StarPU:Aunif}
and hardware \cite{AcceleratorStore}.

\subsection*{Analytical models for the multiprocessor}

Hill and Marty \cite{hill2008amdahl} created an initial analytical
model revealing the relationship between single thread execution speedup
gained from using larger cores, and the multi-threaded phase execution
speedup gained from using more cores, under a total budget. The model
shows that the optimal results are achieved by asymmetric multicores,
with one large core for accelerating the single-threaded phase and
many small cores for accelerating the multi-threaded phases. 

Woo et al. \cite{ExtendingAmda} extended this model for $Performance/Watt$
and $Performance/Joule$ measures. Three systems are modeled: symmetric
full-blown processors, symmetric efficient processors, and asymmetric
ones. The three systems can be compared with one another by limiting
them to an equal power budget. Once again, the asymmetric multicore
shows better results for almost any measure.

Chung et al. \cite{Single-ChipHe} also extended Hill's model. Their
model took into considerations three different budgets: total area
(as in \cite{hill2008amdahl}), total power (similar to \cite{ExtendingAmda}),
and total bandwidth. Three models of chips were tested - symmetric
multicore, asymmetric multicore, and heterogeneous. The last model
is an extension of the asymmetric one, where the efficient cores are
in fact {}``unconventional cores'' and therefore even more power
efficient. The flexibility of unconventional cores was not modeled.
All the unconventional cores in the model can execute the entire parallel
portion of the workload. The authors of the article have made a few
projections on the future technology nodes and the changes in overall
bandwidth budget, and have reached the conclusion that bandwidth,
rather than power, would be the main reason for performance limitation
in the future.

Our technique is different from the ones presented above, by the fact
that we provide optimal solution for $n$ different execution segments
rather than only two segments. Multi-Amdahl models the cost of the
inflexibility introduced in the heterogeneous system.

\section{Multi-Amdahl}

\subsection{Entities}

\begin{figure}[t]
\includegraphics[scale=0.75]{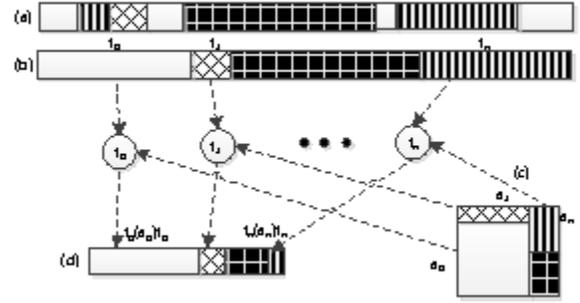}

\caption{\label{fig:basic-model}Basic model (a) BGP execution time, (b) aggregated
BGP execution time, (c) resource allocation (e.g. area), (d) acceleration
function, (e) final execution time}
\end{figure}

Multi-Amdahl is a strategy for finding the optimal resource assignment
for different accelerators sharing common limited resources. 

Figure \ref{fig:basic-model} illustrates an example for the three
basic entities of the technique: 

\textbf{The Workload ---} Figure \ref{fig:basic-model}(a) presents
execution time as measured running on a Basic General-Purpose core
(BGP). Figure \ref{fig:basic-model}(b) presents execution time when
it is aggregated and divided into $n$ segments, each of which will
run on a different accelerator. For simplicity, we do not model any
cost for moving context between segments. This is common in such models
(e.g. \cite{hill2008amdahl,ExtendingAmda,Shee08}). Segment $i$ takes
$t_{i}$ seconds to execute on the BGP. The parameters $t_{i},0\leq i<n$
represent the workload's distribution between the different execution
segments.

\[
T_{BGP}=\sum t_{i}\]

\textbf{Resource and Constraint ---} Figure \ref{fig:basic-model}(c)
illustrates how the chip is divided into $n$ hardware units. The
units share a common resource (e.g. area, power, energy). The chip
design aims at resource allocation under a specific constraint. For
example, when allocating area $A$ to different hardware units, the
constraint is the total die area $A$,

\begin{equation}
\sum x_{i}\leq A\label{eq:constraint-static}\end{equation}
The resource units are normalized so that the BGP uses one unit. Different
resources and constraints are presented in Section \ref{sec:Resource}.

\textbf{Efficiency ---} Figure \ref{fig:basic-model}(d) presents
unit efficiency, which is a function determining how long section
$i$ will take to execute when assigned $x_{i}$ resource.

\[
T_{i}=t_{i}f_{i}(x_{i})\]

Each unit may be described by a different function. The function represents
the unit's technology. For example, when considering number of transistors
(normalized to BGP transistor number) as a resource, the function:

\[
f(x)=\frac{1}{100\sqrt{x}}\]
models an accelerator that is 100 times as efficient as a BGP when
assigned BGP-equivalent transistors, but it will only double its performance
if assigned four times more transistors, according to Pollack\textquoteright{}s
law \cite{Newmicroarchi}. Theses functions hide details of how the
resource is used (e.g. is the area divided into many narrow units
or a few wide ones). 

Figure \ref{fig:basic-model}(e) illustrates the total aggregated
execution time. The optimization goal is to minimize this total time.

\begin{equation}
T_{exec}=\sum T_{i}=\sum t_{i}f_{i}(x_{i})\label{eq:execution-time}\end{equation}
Different use-cases for the model are presented in Section \ref{sec:use-case}.

\subsection{Optimization}

Lagrange multipliers are a mathematical tool for finding maxima and
minima for a multi-dimensional function within a set of constraints
on the input variables. In Multi-Amdahl, we minimize total execution
output (e.g. time) under the constraint imposed by the resource (e.g.
area, power). We assume that additional resources added to the system
will create an output gain (e.g. performance), thus the optimal point
is inside the limited space where the resource budget is exactly met
(i.e., we assume that the Karush-Kuhn-Tucker conditions \cite{NonlinearProgramming}
are met.) Using Equations \eqref{eq:constraint-static} and \eqref{eq:execution-time},
the optimization problem can be formalized as:

\begin{eqnarray*}
\textrm{minimize} & \sum f_{i}(x_{i})t_{i}\\
\textrm{subject to:} & \sum x_{i}=A\end{eqnarray*}

Using the Lagrange optimization method, it follows that the optimal
solution satisfies:

\begin{equation}
f_{i}^{'}(x_{i})t_{i}=f_{j}^{'}(x_{j})t_{j}\label{eq:f'x_t}\end{equation}

The intuition behind Equation \eqref{eq:f'x_t} is that each infinitesimal
additional resource would create the same overall run-time improvement
on any accelerator it would be assigned to. This is the basic equation,
describing static resources. More complex cases can be analyzed in
the same way to provide their optimal point.

\section{\label{sec:Comparing}Comparing with Amdahl's law}

\begin{figure}[t]
\includegraphics[scale=0.45]{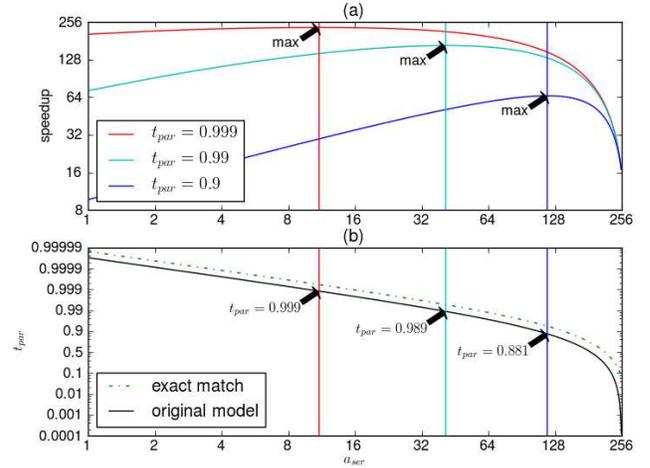}

\caption{\label{fig:us_vs_hill}The serial/parallel problem (a) speedup per
serial CPU size (according to \cite{hill2008amdahl}), (b) optimal
serial CPU size per workload}
\end{figure}

As an example, we will use our optimization technique to implement
a well-known problem of asymmetric processors. This problem is composed
out of two execution segments: parallel and serial.

\[
t_{0}=t_{serial}\quad t_{1}=t_{parallel}\]

The parallel section is run most efficiently on many small cores,
and the serial on a large core. Therefore, the proposed chip should
contain a mixture of both. Hill and Marty \cite{hill2008amdahl} analyzed
the implications of the chip\textquoteright{}s limited area. We will
use area as a resource (i.e. constraint), and normalize total execution
time on the BGP to be 1. 

\begin{eqnarray*}
a_{parallel}+a_{serial} & = & A\\
t_{serial}+t_{parallel} & = & 1\end{eqnarray*}

The speedup of the parallel section is assumed to be proportional
to the number of small cores, and therefore also to the total area
of the small cores, which execute the parallel tasks.

\[
f_{parallel}(a)=\frac{1}{a}\]

The {}``serial accelerator'' is the large CPU, whose performance
scales with area according to Pollack's Law \cite{Newmicroarchi}.
Formalizing this in our model's terms:

\[
f_{serial}(a)=\frac{1}{\sqrt{a}}\]

Applying Equation \eqref{eq:f'x_t} reveals the optimal relation between
the total area of the efficient parallel processors and the area of
the serial processor.

\begin{equation}
a_{parallel}=a_{serial}^{3/4}\sqrt{\frac{2t_{parallel}}{1-t_{parallel}}}\label{eq:a_parallel}\end{equation}

An immediate result of our model is the optimal resource allocation
point. The speedup obtained at this point could be calculated by Equation
\eqref{eq:execution-time}. The focus of this model allows for a different
set of insights. For example, it is apparent from Equation \eqref{eq:a_parallel}
that the serial section grows faster than the parallel one when the
chip receives additional resource. 

Note that for simplicity we presented a model in which the large core
will not be used for execution of the parallel segment. We will use
the assumption that different hardware units do not execute the same
code in following sections, when expanding the model for many accelerators.
On the contrary, Hill and Marty \cite{hill2008amdahl} used a different
assumption, in which the parallel segment is executed by all the cores
on the chip, including the large core. This slightly changes the results,
but the Multi-Amdahl optimization technique does not change significantly.

Figure \ref{fig:us_vs_hill} illustrates the optimization results.
Figure \ref{fig:us_vs_hill}(a), essentially taken from Hill and Marty's
paper \cite{hill2008amdahl}, reveals the existence of optimal resource
allocations ($a_{serial}$) which changes according to the workload
($t_{parallel}$). Figure \ref{fig:us_vs_hill}(b) presents the exact
value of these optimal resource allocations using Multi-Amdahl technique,
and in particular the relation between $t_{parallel}$ and the optimal
value of $a_{serial}.$ Results are presented for both cases, where
the parallel section is executed either only on the efficient cores
or on the entire chip.

\section{\label{sec:Resource}Different resources}

The Multi-Amdahl technique can be applied to different resource types
with various constraints.

\subsection{\label{sub:resource-static}Static resources}

A static resource is used by an accelerator for the entire life-time
of the problem. For example it could be the die area or the number
of transistors. If all hardware units are working concurrently, power
and IO might also be modeled as static resources.

When allocating a static resource, the designer's goal is to stay
within a total budget $X$:

\[
\sum_{i=0}^{n-1}x_{i}\leq X\]

\subsection{The Power Resource}

Today, the resource-allocation efforts of the chip designer have been
shifted to power constraints. When modeling power, we must take into
account both \textit{dynamic power}, which is only consumed when the
unit itself works, and \textit{static power}, which is also consumed
when it is idle.

The actual resource assigned by the designer, however, is still the
number of transistors. Both the static and the dynamic power can be
modeled as proportional to the number of transistors in the accelerator,
so we can model them as linearly dependent. A linear relation between
static and dynamic power is a common model, used e.g. by \cite{ExtendingAmda}.
In our model, each unit is assigned $p_{i}$ power when in use, and
consumes additional $k_{i}p_{i}$ static power all the time. $k_{i}$
is assumed to be another accelerator-technology dependent parameter,
known to the chip manufacturer.

Several constraints can be considered for the power resource:

\textbf{Instantaneous power ---} There is a total power budget that
the multiprocessor may use at any given instant. The budget usually
derives from power dissipation consideration. The constraint is imposed
on each unit (or execution segment) separately. This constraint is
most applicable in case the different segments last long enough to
overheat the chip. 

\[
\forall0\leq i<n:\quad p_{i}+\sum_{j=0}^{n-1}k_{j}p_{j}\leq P\]

\textbf{Energy ---} The energy represents the total power consumed
by the chip over time. It is a design goal for servers, where electricity
costs are considerable, and for mobile devices, where minimizing energy
consumption is necessary to maximize battery life. 

\[
\sum k_{j}p_{j}\sum f_{i}(p_{i})t_{i}+\sum f_{i}(p_{i})t_{i}p_{i}\leq E\]

\textbf{Total Dynamic Power ---} The total (or average) dynamic power
is calculated by dividing the overall energy by the overall execution
time. If execution segments are short enough, power dissipation poses
a constraint on the average power consumption, rather than on the
instantaneous one.

\[
\sum k_{i}p_{i}+\frac{\sum f_{i}(p_{i})t_{i}p_{i}}{\sum f_{i}(p_{i})t_{i}}\leq TDP\]

\subsection{Multiple resources}

More than one constraint and resource can be tested at the same time.
For example, we are going to discuss the combined effect of assigning
supply voltage (marked $v_{i}$) and area ($a_{i}$).

The maximum operation frequency is proportional to voltage. 

\[
freq_{i}=v_{i}\]

Performance is modeled as linearly proportional to frequency, and
sub-linearly proportional to area:

\[
f_{i}(a_{i},v_{i})=\frac{1}{freq_{i}\sqrt{a_{i}}}=\frac{1}{v_{i}\sqrt{a_{i}}}\]

Energy is modeled as proportional to area, voltage, and operation
time:

\[
E_{i}=f_{i}(a_{i},v_{i})t_{i}a_{i}v_{i}^{3}\]

We have two constraints: one for total area, and one for total energy.

\[
\sum a_{i}\leq A\]

\[
\sum f_{i}(a_{i},v_{i})t_{i}a_{i}v_{i}^{3}\leq E\]

Note that no constraints are directly applied to the voltage. 

\textbf{}

\section{\label{sec:use-case}Different use-cases}

Multi-Amdahl could be used to describe different use cases. A use
case determines the workload and efficiency functions.

\subsection{Serial Execution}

One case of introducing accelerators into a system is when little
or no parallelism can be extracted from the code. In this simplified
model, we assume the entire chip only executes one segment at a time,
and only on the appropriate accelerator. The total execution time
is given by:

\[
T_{exec}=\sum T_{i}\quad T_{i}=t_{i}f_{i}(x_{i})\]
This was the use-case of all the previous sections.

\subsection{\label{sub:use-ser}Parallel Execution}

In this model, the various accelerators handle a different type of
parallel input each. We try to minimize the average latency, given
by:

\[
T_{latency}=\sum\lambda_{i}f_{i}(x_{i})\]

$\lambda_{i}$ is the rate (inputs per second) for this type of input.

$f_{i}$ is the latency of calculation of type $i$ when assigned
$x_{i}$ resource.

This model can be applied, e.g. for network processors, where different
accelerators handle different types of packages (such as encrypted,
compacted..).

\subsection{Optimizing for different units in a CPU}

Even inside a basic CPU there are various separate hardware units,
some of which can be described as handling their own instruction set.
For example, we might consider allocating resource optimally between
3 units: the cache, the branch predictor, and the ALU.

\[
CPI=\lambda_{c}c_{c}(x_{c})+\lambda_{p}c_{p}(x_{p})+\lambda_{a}c_{a}(x_{a})\]

$\lambda_{c}$ is the number of memory accesses per instructions.
In our model, memory accesses are executed by the cache. As the cache
is assigned more resource, it becomes larger and the cache hit ratio
increases.

\[
c_{c}(x)=hit\%(x)*T_{hit}+(1-hit\%(x))*T_{miss}\]

$\lambda_{p}$ is the number of branches per instruction. Branches
are modeled as executed by the branch predictor. As the cache predictor
is assigned more resource, it should improve branch prediction rates.

\[
c_{p}(x)=(1-predict\%(x))*T_{mispredict}\]

$\lambda_{a}$ is the number of ALU instructions. As ALU is assigned
more resource, more ALUs are added to the system which increases throughput
of ALU instructions. 

\[
c_{a}(x)=\frac{T_{ALU}}{x}\]

\section{Initial Results}

Initial results were calculated for a static resource, such as area,
and for the serial execution model. Our results indicate that good
accelerator efficiency can be put to use even for the price of flexibility.
With dynamic resource, the cost of inflexibility is much smaller,
and in that sense our results are conservative.

\subsection{The general case}

When allocating resources for different accelerators, we must take
two elements into consideration: how efficient the accelerator is
and how useful it is (meaning, what is its part in the workload).
We consider a general efficiency function:\[
f_{i}(a)=\frac{1}{\alpha_{i}a^{\beta_{i}}}\]

A general-purpose CPU can be modeled by $\alpha_{0}=1$. Higher values
of $\alpha$ are assigned to more efficient accelerators. We use Multi-Amdahl
to extract the appropriate area allocation for each of the accelerators:\begin{equation}
a_{i}=a_{0}^{\frac{\beta_{j}+1}{\beta_{i}+1}}\left(\frac{\alpha_{0}/\beta_{0}}{\alpha_{i}/\beta_{i}}\frac{t_{i}}{t_{0}}\right)^{\frac{1}{\beta_{i}+1}}\label{eq:area_general_result}\end{equation}

The most interesting thing this solution reveals is that the workload-dependent
parameters ($t_{i}$) have equal or lower importance to the parameters
that are dependent on the accelerator's technology ($\alpha_{i}$,
$\beta_{i}$) when determining the optimal solution. This has an implication
on the chip manufacturer's ability to allocate resources properly
with even a partial knowledge of the workload, which will be analyzed
later.

\subsection{Effective heterogeneous speedup}

A heterogeneous system might consist of various units, when each can
accelerate its designated code segment with noticeable speedup over
a general-purpose machine. The program is composed of various segments.
The effective speedup is measured over the entire execution, including
the general-purpose section, and is generally lower then the speedup
for a single section. Multi-Amdahl reveals another effect of heterogeneity.
As more heterogeneity is added to the system, the resource is shared
between more accelerators, and therefore each accelerator is assigned
less resource, thus reducing its speedup.

To display this effect, we consider a system composed of one general-purpose
section and $n$ accelerators (notice this system has $n+1$ segments).
All code, accelerated or not, is assumed to be entirely parallelisable
($\beta_{i}=1$). All accelerators are equally efficient ($\alpha_{i>0}=\alpha$).

\[
f_{0}(a)=\frac{1}{a}\quad f_{i>0}(a)=\frac{1}{\alpha a}\]

We mark $\delta$ to be the part of the original code using the accelerators.
We assume this part is equally distributed among the different accelerated
segments.

\[
t_{0}=1-\delta\quad t_{i>0}=\frac{\delta}{n}\]
Putting this into Equation \eqref{eq:area_general_result}:

\[
a_{i}=a_{0}\sqrt{\frac{\delta}{\alpha n(1-\delta)}}\]
from which we can also derive the total execution time for a chip
with area budget $A$, using Equation \eqref{eq:execution-time}:

\[
T_{het}=\frac{1}{A}\left(2\sqrt{\frac{n}{\alpha}\delta(1-\delta)}+1-\delta\left(1-\frac{n}{\alpha}\right)\right)\]

A homogeneous multicore system uses the entire available area for
general-purpose CPU, and executes the entire code without any speedup:

\[
T_{hom}=\frac{1}{A}\]
Therefore, the speedup from introducing heterogeneity into the system
is:

\[
Speedup_{het}=\frac{T_{hom}}{T_{het}}=\left(2\sqrt{\frac{n}{\alpha}\delta(1-\delta)}+1-\delta\left(1-\frac{n}{\alpha}\right)\right)^{-1}\]

\begin{figure}[t]
\includegraphics[scale=0.5]{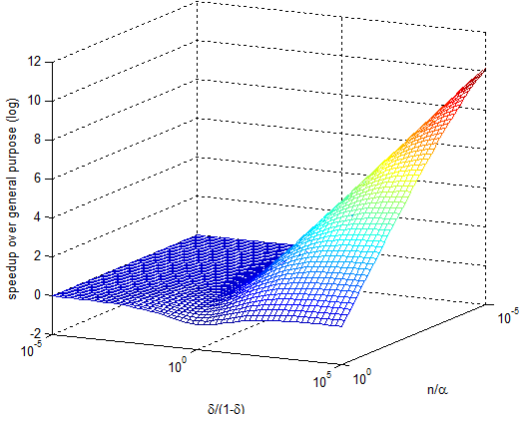}

\caption{\label{fig:n_a_de}Combined effect of $\frac{n}{\alpha}$and $\delta$
on speedup}
\end{figure}

\textbf{Speedup vs. Flexibility ---} The value of $n$ has a tremendous
influence on the speedup gained. This is the effect of accelerator's
inflexibility. Accelerators use system resources all the time, but
they are seldom used for actual computation. The more accelerators
are in the multiprocessor, the more {}``dead area'' it contains,
per execution segment.

One reason for adding accelerators into a system would be to create
more specific, and therefore more efficient, accelerators. The equations
present a linear relation between $n$ and $\alpha$, which fits intuition.
An accelerator capable of two operations should be split into two
accelerators capable of one operation each, only if those two accelerators
are at least twice as efficient as the previous one.

\textbf{Speedup vs. Code coverage --- }Another reason for adding accelerators
to an existing system would be moving part of the code that previously
ran on the general processor to an accelerator, namely to increase
$\delta$.

Figure \ref{fig:n_a_de} reveals that for low values of $\delta$
there is little effect either way. As $\delta$ approaches $1$, the
rule of thumb is that multiplying the number of accelerators is worthwhile
if it does better than halve the amount of code not running on accelerators
($1-\delta$).

\begin{figure}[t]
\includegraphics[scale=0.5]{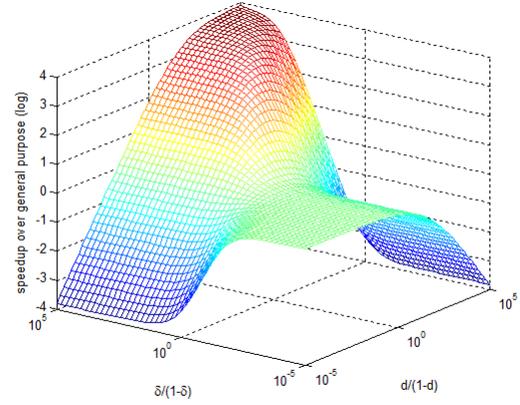}

\caption{\label{fig:sensitivity}Sensitivity $(\frac{n}{\alpha}=\frac{1}{50})$}
\end{figure}

\subsection{Resource allocation sensitivity}

As we have previously mentioned, the chip manufacturer is unaware
of the exact nature of the workload running on the machine, and can
only estimate the expected workloads. For that reason we also model
the case where the manufacturer creates a chip for a given workload,
while the actual workload is different.

In our case, the chip manufacturer assumes equal use of all accelerators,
and assumes use of accelerated code $\delta=d$. According to these
assumptions, the manufacturer divides the area between the accelerators.

\[
a_{i}=a_{0}\sqrt{\frac{d}{\alpha n(1-d)}}\]

However, the actual value of $\delta$ is different. The speedup of
the chip for the actual workload will be:

\[
Speedup=\left(\left(1+\frac{\delta}{d}-2\delta\right)\sqrt{\frac{n}{\alpha}\frac{d}{(1-d)}}+1-\delta\left(1-\frac{n}{\alpha}\right)\right)^{-1}\]

Figure \ref{fig:sensitivity} presents a few properties of the equation:
There is no reason for the chip manufacturer to assume small values
for $d$ (significantly smaller then $0.5\mbox{)}$. For those values,
high accelerator use will incur a serious slowdown, as most execution
time is spent on an accelerator with insufficient resource, while
low accelerator use results in minimal speedup, because most of the
execution is on the CPU. Large values of $d$ (close to $1$) might
prove very beneficial for enough accelerator usage, but will be destructive
if the accelerators are not used, as the CPU is very weak. Using $0.5$,
or somewhat larger values for $d$ is the {}``safe'' choice. An
observable speedup can be seen when the accelerator is used, while
the slowdown for workloads not using the accelerator is negligible. 

\textbf{}

\section{CONCLUSION AND FUTURE WORK }

This paper presents Multi-Amdahl, an analytical technique for optimal
resource allocation in a heterogeneous chip. Our technique relies
on the modeling of the resource, the workload, and the accelerators\textquoteright{}
performance as a function of the chip\textquoteright{}s resource.
We have shown the technique's applicability to a large field of problems.
For example, the accelerators considered may either be part of the
general-purpose cores or separate accelerators. 

We have used our model to test the importance of accelerator efficiency
vs. code coverage, and have found the two parameters to be equal when
looking for the optimal resource allocation. We have also discussed
the case of workload variance, and found a {}``sweet-spot'' for
chip design, characterized by minimal slowdown when the accelerator
is not used, versus a measurable speedup when the accelerator is used.
Those results are based on an environment in which moving context
between accelerators has no overhead, and resource allocation is static. 

Generally speaking, our results suggest that inflexibility is a reasonable
price to pay for efficiency, and that accelerator-based heterogeneous
multicores are a promising direction for future chip architectures.

Our future research will concentrate on the expansion of the applications
area, while putting more overhead and resource allocations constrains.

\section*{Acknowledgment}

This research work was partly supported by an Intel research grant
on heterogeneous computing, by the European Research Council Starting
Grant n$^{\circ}$ 210389, by the Hasso Plattner Center for Scalable
Computing, and by the Israeli MOST CMP Research Center. 

\bibliographystyle{IEEEtran}
\addcontentsline{toc}{section}{\refname}\bibliography{IEEEabrv,hetero_bib}

\end{document}